\documentstyle[12pt,epsf]{article}

\begin{document}

\baselineskip 6mm
\renewcommand{\thefootnote}{\fnsymbol{footnote}}


\newcommand{\nc}{\newcommand}
\newcommand{\rnc}{\renewcommand}


\rnc{\baselinestretch}{1.24}    
\setlength{\jot}{6pt}       
\rnc{\arraystretch}{1.24}       

\makeatletter
\rnc{\theequation}{\thesection.\arabic{equation}}
\@addtoreset{equation}{section}
\makeatother



\nc{\be}{\begin{equation}}
\nc{\ee}{\end{equation}}
\nc{\bea}{\begin{eqnarray}}
\nc{\eea}{\end{eqnarray}}
\nc{\xx}{\nonumber\\}
\nc{\ct}{\cite}
\nc{\la}{\label}
\nc{\eq}[1]{(\ref{#1})}
\nc{\newcaption}[1]{\centerline{\parbox{6in}{\caption{#1}}}}

\nc{\fig}[3]{
\begin{figure}
\centerline{\epsfxsize=#1\epsfbox{#2.eps}}
\newcaption{#3. \label{#2}}
\end{figure}
}


\def\CA{{\cal A}}
\def\CC{{\cal C}}
\def\CD{{\cal D}}
\def\CE{{\cal E}}
\def\CF{{\cal F}}
\def\CG{{\cal G}}
\def\CH{{\cal H}}
\def\CL{{\cal L}}
\def\CM{{\cal M}}
\def\CN{{\cal N}}
\def\CO{{\cal O}}
\def\CP{{\cal P}}
\def\CS{{\cal S}}
\def\CV{{\cal V}}
\def\CW{{\cal W}}
\def\CY{{\cal Y}}


\def\IR{{\hbox{{\rm I}\kern-.2em\hbox{\rm R}}}}
\def\IB{{\hbox{{\rm I}\kern-.2em\hbox{\rm B}}}}
\def\IN{{\hbox{{\rm I}\kern-.2em\hbox{\rm N}}}}
\def\IC{\,\,{\hbox{{\rm I}\kern-.59em\hbox{\bf C}}}}
\def\IZ{{\hbox{{\rm Z}\kern-.4em\hbox{\rm Z}}}}
\def\IP{{\hbox{{\rm I}\kern-.2em\hbox{\rm P}}}}
\def\IH{{\hbox{{\rm I}\kern-.4em\hbox{\rm H}}}}
\def\ID{{\hbox{{\rm I}\kern-.2em\hbox{\rm D}}}}


\def\a{\alpha}
\def\b{\beta}
\def\ga{\gamma}
\def\d{\delta}
\def\ep{\epsilon}
\def\ph{\phi}
\def\k{\kappa}
\def\l{\lambda}
\def\m{\mu}
\def\n{\nu}
\def\th{\theta}
\def\rh{\rho}
\def\s{\sigma}
\def\t{\tau}
\def\w{\omega}
\def\G{\Gamma}


\def\vare{\varepsilon}
\def\bz{\bar{z}}
\def\bw{\bar{w}}


\def\half{\frac{1}{2}}
\def\dint#1#2{\int\limits_{#1}^{#2}}
\def\goto{\rightarrow}
\def\para{\parallel}
\def\brac#1{\langle #1 \rangle}
\def\grad{\nabla}
\def\curl{\nabla\times}
\def\div{\nabla\cdot}
\def\p{\partial}
\def\e{\epsilon_0}


\def\Tr{{\rm Tr}\,}
\def\det{{\rm dot}}


\def\cp{{\bf CP}^n}
\def\s{{\bf S}}
\def\r{{\bf R}}
\def\c{{\bf C}}
\def\z{{\bf Z}}

\begin{titlepage}
\hfill\parbox{4cm} {{\tt hep-th/0105087}} \vspace{25mm}

\begin{center}
{\Large \bf $\cp$ Model on Fuzzy Sphere} \\
\vspace{15mm}
Chuan-Tsung Chan$^a$\footnote{ctchan@ep.nctu.edu.tw}, Chiang-Mei
Chen$^b$\footnote{cmchen@phys.ntu.edu.tw}, Feng-Li
Lin$^c$\footnote{linfl@phya.snu.ac.kr}, and Hyun Seok
Yang$^b$\footnote{hsyang@phys.ntu.edu.tw}
\\[10mm]
$^a${\sl Department of Electrophysics, National Chiao Tung University \\
Hsinchu 300, Taiwan, R.O.C.}\\
$^b${\sl Department of Physics, National Taiwan University \\
Taipei 106, Taiwan, R.O.C.}\\
$^c${\sl School of Physics \& Center for Theoretical Physics \\
Seoul National University, Seoul 151-742, Korea}
\end{center}

\thispagestyle{empty} \vskip 20mm

\centerline{\bf ABSTRACT}
\vskip 4mm
\noindent

We construct the $\cp$ model on fuzzy sphere. The Bogomolny bound
is saturated by (anti-)self-dual solitons and the general
solutions of BPS equation are constructed. The dimension of moduli
space describing the BPS solution on fuzzy sphere is exactly the
same as that of the commutative sphere or the (noncommutative)
plane. We show that in the soliton backgrounds, the number of
zero modes of Dirac operator on fuzzy sphere, Atiyah-Singer
index, is exactly given by the topological charge of the
background solitons.

\vspace{2cm}
\end{titlepage}

\baselineskip 7mm
\renewcommand{\thefootnote}{\arabic{footnote}}
\setcounter{footnote}{0}

\section{Introduction}

A noncommutative space is obtained by quantizing a given space
with its symplectic structure, treating it as a phase space. Also
field theories can be formulated on a noncommutative space.
Noncommutative field theory means that fields are defined as
functions over noncommutative spaces. At the algebraic level, the
fields become operators acting on a Hilbert space as a
representation space of the noncommutative space. Since the
noncommutative space resembles a quantized phase space, the idea
of localization in ordinary field theory is lost. The notion of a
point is replaced by that of a state in representation space.

Quantum field theory on a noncommutative space has been proved to
be useful in understanding various physical phenomena, like as
various limits of M(atrix) theory compactification \ct{cds,ho},
low energy effective field theory of D-branes with constant
Neveu-Schwarz $B$-field background \ct{aas,sw}, and quantum Hall
effect \ct{bell}. Although noncommutative field theories are
non-local, they appear to be highly constrained deformation of
local field theory. Thus it may help understanding non-locality at
short distances in quantum gravity.

The fuzzy sphere is constructed by introducing a cut-off parameter
$N$ for angular momentum of the spherical harmonics: $\{{\hat
Y}_{lm}; l\leq N\}$ \ct{Madore}. Thus the number of independent
functions is $\sum_{l=0}^N (2l+1) = (N+1)^2$. In order for this
set of functions to form a closed algebra, the functions are
replaced by $(N+1) \times (N+1)$ hermitian matrices and then the
algebra on the fuzzy sphere is closed \ct{Hoppe}. Consequently,
the algebra on the fuzzy sphere becomes noncommutative. The
commutative sphere is recovered for $N\rightarrow\infty$. One of
the attractive features of the fuzzy sphere is that it is
covariant with respect to $SO(3)$ like the commutative sphere.

Recently, it has been shown that the fuzzy sphere is a natural
candidate for the quantum geometry due to stringy effects in the
AdS/CFT duality \ct{holi} and that the field theories on fuzzy
sphere appear naturally from D-brane world-volume theory
\ct{ars,WZW} and matrix theory with some backgrounds \ct{iktw}.
Interestingly, it was argued based on the $SU(2)$ WZW model that
the RR charges of spherical D2-branes are only defined modulo
some integer \ct{WZW},  which are $U(1)$ charges defined on
D2-brane world-volume (fuzzy sphere). This was confirmed using
K-theory calculation in \ct{K-theory}. Many efforts to construct
field theories on the fuzzy sphere were also pursued in
\ct{ftfs,GKP,Pres}.

In this paper we construct the $\cp$ model on fuzzy sphere. In
section 2, the fuzzy sphere is constructed by using the
noncommutative version of Hopf fibration $\pi:\s^3 \rightarrow
\s^2$, which is essentially based on the Holstein-Primakoff
realization of $SU(2)$ algebra \ct{HP}. Based on this
realization, the derivative operators on fuzzy sphere are
defined. In section 3, the $\cp$ model on fuzzy sphere is
constructed. Our present construction of $\cp$ model closely
follows that of Berg and L\"uscher \ct{BL}, in fact, the
noncommutative generalization of them. It is shown in section 4
that the Bogomolny bound is saturated by (anti-)self-dual
solitons and the general solutions of BPS equation are
constructed. The dimension of moduli space describing the BPS
solution on fuzzy sphere is exactly the same as that of the
commutative sphere \ct{BL} or the (noncommutative) plane
\ct{LLY}. In section 5 we show that in the soliton backgrounds,
the Atiyah-Singer index, that is the number of zero modes of
Dirac operator on fuzzy sphere, is exactly given by the
topological charge of the background solitons. In section 6 we
address a topological issue on the BPS solitons on fuzzy sphere
and some other issues related to our work. In Appendix, we
explain the fuzzy spherical harmonics ${\hat Y}_{lm}$, the
Clebsch-Gordan decomposition of tensor products, and the Casimir
operator of $SU(2)$.

\section{Fuzzy Sphere from Hopf Fibration}

The algebra of the fuzzy sphere \ct{Madore} is generated by $\hat
r_a$ satisfying the commutation relations
\begin{equation}\label{NCR}
[ \hat r_a, \, \hat r_b ] = i \alpha\, \epsilon_{abc}\, \hat r_c,
\qquad (a,b,c=1,2,3)
\end{equation}
as well as the following condition for $\hat r_a$:
\begin{equation}\la{R2}
\hat r_a \hat r_a = R^2.
\end{equation}
The noncommutative coordinates of (\ref{NCR}) can be represented
by the generators of the $(N+1)$-dimensional irreducible
representation of $SU(2)$
\begin{equation}\la{rL}
\hat r_a = \alpha \hat L_a,
\end{equation}
where
\begin{equation}
[ \hat L_a, \, \hat L_b ] = i \epsilon_{abc}\, \hat L_c.
\label{su2}
\end{equation}
Since the  second Casimir of $SU(2)$ in the $(N+1)$-dimensional
irreducible representation is given by $N(N+2)/4$, thus $\alpha$
and $R$ are related by the following relation
\begin{equation}
R^2 = \alpha^2 \frac{N(N+2)}4.
\end{equation}
In the $\alpha \to 0$ limit, $\hat r_a$ describe commutative
sphere:
\begin{equation}
r_1 = R \sin\theta \cos\phi, \quad r_2 = R \sin\theta \sin\phi,
\quad r_3 = R \cos\theta.
\end{equation}

Since $\s^2$ is not parallelizable unlike $\s^3\simeq SU(2)$, the
module of derivations on $\s^2$ is not free \ct{Madore}. If we
enlarge the coordinate space from $\s^2$ to $\s^3$ by the addition
of a $U(1)$ gauge degree of freedom, we can have a free module of
the derivations (acting on $\s^3$). This is a well-known
construction, called the Hopf fibration of $\s^2$. Indeed $\s^3$
can be regarded as a principal fiber bundle with base space $\s^2$
and a $U(1)$ structure group. Equivalently,
\begin{equation}\label{coset}
  \s^2 \simeq SU(2)/U(1),
\end{equation}
where $U(1)$ is the subgroup of $SU(2)$. A complex scalar field on
$\s^2$ can then be identified with a smooth section of this
bundle.

The Hopf fibration $\pi:\s^3 \rightarrow \s^2$ can be generalized
to the noncommutative space ${\bf C}^2$ satisfying the relations
\begin{equation} \la{bi-harm}
[a_\a, a_\b] = [a_\a^\dag, a_\b^\dag] = 0, \qquad [a_\a,
a_\b^\dag] = \delta_{\a\b},\qquad (\a,\b,=1,2)
\end{equation}
as follows
\begin{equation} \la{Hopf}
\hat L_a = \frac12 \xi^\dag \sigma_a \xi, \qquad \xi = \left(
\begin{array}{c} a_1 \\ a_2 \end{array} \right),
\la{gens}
\end{equation}
where $\sigma_a$ are the Pauli matrices and $\xi$ is an $SU(2)$
spinor with the normalization $\xi^\dag \xi=N$. (Based on this
Hopf fibration, topologically nontrivial field configurations on
fuzzy sphere were discussed in \ct{GKP,Pres}.) It is
straightforward to check for $\hat L_a$'s of \eq{gens} to satisfy
the $SU(2)$ algebra \eq{su2}. Now the $SU(2)$ generators are
given by $\c^2$ coordinates as
\begin{equation}\la{Ls}
\hat L_1 = \frac12 (a_1 a_2^\dag + a_1^\dag a_2), \quad \hat L_2 =
\frac{i}2 (a_1 a_2^\dag - a_1^\dag a_2), \quad \hat L_3 = \frac12
(a_1^\dag a_1 - a_2^\dag a_2),
\end{equation}
which is, in fact, the Schwinger realization of $SU(2)$ algebra.
The associated ladder operators are defined as
\begin{equation}
\hat L_+ = \hat L_1 + i \hat L_2 = a_1^\dag a_2, \quad \hat L_- =
\hat L_1 - i \hat L_2 = a_1 a_2^\dag
\end{equation}
and their communication relations are
\begin{equation}
[\hat L_+, \hat L_-] = 2 \hat L_3, \qquad [\hat L_3, \hat L_\pm] =
\pm \hat L_\pm.
\end{equation}
Note that the $SU(2)$ generators in \eq{gens} are invariant under
the transformation
\begin{equation}\label{U1}
  \xi \to e^{i \psi} \xi, \quad
  \xi^\dag \to \xi^\dag e^{-i \psi},
\end{equation}
showing that the fiber is $U(1)$.

The $(N+1)$-dimensional irreducible representation of $SU(2)$,
denoted as ${\cal H}_N$, can be given by the following orthonormal
basis
\begin{equation}\label{basis}
|n\rangle =|\frac{N}2, n-\frac{N}2\rangle ={ (a_1^\dag)^{n}
(a_2^\dag)^{N-n}\over \sqrt{n!(N-n)!}}|0\rangle_{12}, \qquad
(n=0,1,\cdots, N),
\end{equation}
where $|j, m\rangle\;(j \in \z/2)$ is a spherical harmonics and
$|0\rangle_{12}$ is the vacuum defined by
$a_1|0\rangle_{12}=a_2|0\rangle_{12}=0$. Let ${\cal A}_N$ be
operator algebra acting on the $(N+1)$-dimensional Hilbert space
${\cal H}_N$, which can be identified with the algebra Mat($N+1$)
of the complex $(N+1) \times (N+1)$ matrices. Then the
integration over the fuzzy sphere is given by the trace over
${\cal H}_N$ and defined by \footnote{In the commutative limit,
$\Tr$ over matrices is mapped to the integration over functions
as $\Tr \rightarrow \int \frac{d\Omega}{4\pi}$.}
\begin{equation}\label{trace}
\Tr {\cal O} = \frac1{N+1} \sum_{n=0}^N < n | {\cal O} | n >,
\end{equation}
where ${\cal O} \in {\cal A}_N$.

Let's consider a scalar field $\Phi$ on the noncommutative $\c^2$
defined by \eq{bi-harm} (or on $\s^3$ after the restriction
$\xi^\dag \xi=N$) of the form \ct{GKP}
\begin{equation}\label{Phi}
\Phi=\sum \Phi_{m_1m_2n_1n_2}{a_1^\dag}{}^{m_1}{a_2^\dag}{}^{m_2}
a_1^{n_1}a_2^{n_2}.
\end{equation}
The above scalar field $\Phi$ can be classified according to the
$U(1)$ gauge transformation \eq{U1} and the set of fields $\Phi$
with definite $U(1)$ charge $k$ will be denoted as $\Phi_k$:
\begin{equation}\label{U1tf}
  \Phi_k \;\;\to \;\;\Phi_k e^{-ik\psi},
\end{equation}
where $k=m_1 +m_2 -n_1 -n_2 \in {\bf Z}$. Indeed the field $\Phi$
is a section of the $U(1)$ bundle over $\s^2$ with a definite
value $k$ along the fiber and the number $k$ labels the
equivalence classes (homotopy classes) of $\Phi$ in \eq{Phi}
according to the Hopf fibration \eq{coset}. In section 4, we will
show that this number is related to the topological charge of
$\cp$ solitons, $|Q|=k$.

If we define an operator denoted as $\hat K_3$ for later use
\begin{equation}\label{K3}
\hat K_3 = -N + (a_1^\dag a_1 + a_2^\dag a_2),
\end{equation}
we see that $\Phi_k$ in \eq{U1tf} is an eigenfunction of this
operator, namely
\begin{equation}\label{K3=k}
[\hat K_3, \Phi_k]=k \Phi_k.
\end{equation}
By the analogy from \ct{BL}, $\hat K_3$ may be identified with a
derivation along the fiber, i.e. a Killing vector along $U(1)
\subset SU(2)$. For this reason we need another two derivations
(tangent to $\s^2 \subset SU(2)$) to form a closed $SU(2)$
algebra. There is a unique (up to sign) choice on the generators,
which is essentially based on the Holstein-Primakoff realization
of $SU(2)$ algebra \ct{HP} for two atomic spins, and they are
given by
\begin{equation}\la{K+-}
\hat K_+ = a_1^\dag \sqrt{N- a_1^\dag a_1} +a_2^\dag \sqrt{N-
a_2^\dag a_2}, \quad \hat K_- = K_+^\dag,
\end{equation}
where
\begin{equation}
\hat K_{\pm} = \hat K_1 \pm i \hat K_2.
\end{equation}
It is straightforward to check the $SU(2)$ algebra
\begin{equation}
[ \hat K_a, \, \hat K_b ] = i \epsilon_{abc}\, \hat K_c,
\label{Ksu2}
\end{equation}
or
\begin{equation}
[\hat K_+, \hat K_-] = 2 \hat K_3, \qquad [\hat K_3, \hat K_\pm] =
\pm \hat K_\pm.
\end{equation}
Then the derivatives of an operator ${\cal O}$ are defined by the
adjoint action of $\hat K_a$: \footnote{Since the coordinates of
fuzzy sphere (or $\s^3$) are Lie algebra elements, the derivatives
on the fuzzy sphere (or $SU(2)$ manifold) can be defined as usual
as an endomorphism or an adjoint operation of Lie algebra like as
\eq{deriv} \ct{Jacob}. Then the Leibnitz rule should be obvious in
this definition: $\hat \nabla_a ({\cal O}_1 {\cal O}_2)=(\hat
\nabla_a {\cal O}_1){\cal O}_2+ {\cal O}_1 (\hat \nabla_a {\cal
O}_2)$.}
\begin{equation} \label{deriv}
\hat \nabla_a {\cal O} = i[\hat K_a,\, {\cal O}].
\end{equation}
Thus the generators $\hat K_a$ are the derivations acting on the
Hopf bundle \eq{coset}. \footnote{The reason $\hat{L}_a$'s are
not chosen as the derivations is that, for the fibration
\eq{coset}, $\hat L_a$'s are not proper derivative to expose
topologically nontrivial field configurations since $\hat L_a:
\Phi_k \to \Phi_{k}$. In this sense, $\hat K_a$'s are more
appropriate for our problem since they allow us to directly
separate the $U(1)$ symmetry \eq{U1} from $SU(2)$.}

As pointed out in \ct{GKP} (also in \ct{BL}), for the description
of topologically nontrivial field configurations (with $k\neq 0$),
the field $\Phi$ in \eq{Phi} needs to be a mapping from $\CH_N$ to
$\CH_M$ where $M-N \neq 0$. For $\Phi_k$ in \eq{U1tf}, $M-N=k$.
However, note that, although $\hat K_{\pm}: \Phi_k \to \Phi_{k \pm
1}$ and $\Phi_k: \CH_N \to \CH_{N+k}$, the relevant quantities
such as the action and topological charge densities are the
operations keeping the representation space: $\CH_N \to \CH_{N}$,
so the trace \eq{trace} of these densities is well-defined.

\section{$\cp$ Model on Fuzzy Sphere}

The $\cp$ model manifold is defined by an $(n+1)$-dimensional
complex vector $\Phi=(\phi_1, \phi_2, \cdots, \phi_{n+1})$ of unit
length with the equivalence relation under the overall phase
rotation $\Phi \sim e^{i\theta} \Phi$ \ct{gpe,ddl,Witten}. This
complex projective space of real dimensions $2n$ is equivalent of
the coset space $U(n+1)/U(1)\times U(n)$.

For the purpose of manifestly $SU(2)$ invariant action, we will
not impose the condition \eq{K3=k} for a moment. Later we will
describe how to naturally project the theory onto the fuzzy sphere
by the restriction \eq{K3=k} which specifies the homotopy class of
$\Phi$ in terms of $U(1)$ fiber.

Since the derivation of the field variable $\Phi(\hat r_a)$ is
given by the adjoint action of $\hat K_a$ as in (\ref{deriv}), the
natural action for the $\cp$ model turns out to be
\begin{equation} \label{L1}
S = \Tr \left[ (\hat \nabla_a \Phi)^\dag (\hat \nabla_a \Phi) +
(\Phi^\dag \hat \nabla_a \Phi) (\Phi^\dag \hat \nabla_a \Phi)
\right],
\end{equation}
with the constraint
\begin{equation} \label{Cons}
\Phi^\dag \Phi = 1.
\end{equation}
Precisely speaking, the normalization in \eq{Cons} means that
$\Phi^\dag \Phi$ is an identity operator acting on the
representation space $\CH_N$. This theory has a global $U(n+1)$
symmetry and a local $U(1)$ symmetry
\begin{equation} \label{GT1}
\Phi(\hat r) \to \Phi(\hat r) g(\hat r), \quad g(\hat r) \in U(1),
\end{equation}
which removes the degrees of freedom for an overall $U(1)$ phase
of $\Phi$. The $U(1)$ gauge transformation acts on the right hand
side, which leaves the constraint (\ref{Cons}) invariant. This
ordering of the gauge transformation is the key point which makes
the whole theory work \ct{LLY}.

The above action with the constraint (\ref{Cons}) can be rewritten
as
\begin{equation} \label{L2}
S = \Tr \left[ (\hat D_a \Phi)^\dag (\hat D_a \Phi) + \lambda
(\Phi^\dag \Phi - 1) \right],
\end{equation}
with
\begin{equation}
\hat D_a \Phi = \hat \nabla_a \Phi - i \Phi \hat A_a,
\end{equation}
where $\hat A_a(\hat r)$ is the $U(1)$ gauge field without kinetic
term and $\lambda(\hat r)$ is a Lagrange multiplier to incorporate
the constraint (\ref{Cons}). As there are no derivatives of $\hat
A_a$, one can solve the $\hat A_a$ equation to get
\begin{equation}\la{A}
\hat A_a = -i \Phi^\dag \hat \nabla_a \Phi.
\end{equation}
Thus $\Phi^\dag \hat D_a \Phi$=0. Note that $\hat D_3 \Phi =0$ if
we require the condition \eq{K3=k}.

This action is invariant under the local gauge transformation
defined by (\ref{GT1}) and
\begin{equation}
\hat A_a \to g^\dag \hat A_a g - i g^\dag \hat \nabla_a g.
\end{equation}
Since the field strength of the gauge field $\hat A_a$ is defined
as the curvature tensor of the covariant derivative $\hat D$ by
$[\hat D_a,\, \hat D_b]\Phi =- i \Phi \hat F_{ab}-\epsilon_{abc}
\hat D_c \Phi$, then one can find
\begin{eqnarray}\la{F}
\hat F_{ab} &=& \hat \nabla_a \hat A_b - \hat \nabla_b \hat A_a +
i [\hat A_a,\, \hat A_b] + \epsilon_{abc} \hat A_c
\nonumber\\
&=&-i\Bigl[(\hat D_a \Phi)^\dag \hat D_b \Phi-(\hat D_b \Phi)^\dag
\hat D_a \Phi \Bigr].
\end{eqnarray}

In order to solve the constraint \eq{Cons}, it is convenient to
parameterize the field as follows
\begin{equation} \la{W}
\Phi = W \frac1{\sqrt{W^\dag W}},
\end{equation}
where $W$ is an $(n+1)$-dimensional vector. We also introduce an
$(n+1)$-dimensional projection operator
\begin{equation}\la{P}
P \equiv 1 - \Phi \Phi^\dag = 1 - W \frac1{W^\dag W} W^\dag,
\end{equation}
whose kernel is a one-dimensional space generated by $W$ vector.
In terms of these field variables, the action (\ref{L2}) becomes
\begin{equation}\la{Waction}
S = \Tr \left( \frac1{W^\dag W} \hat \nabla_a W^\dag P \hat
\nabla_a W \right).
\end{equation}
Then, one can check that the above action has a local scaling
symmetry, $W \to W \Delta(\hat r)$, as on the (non-)commutative
plane \ct{LLY}. In addition, there is a still local $U(1)$ gauge
symmetry $W \to W e^{i\Lambda(\hat r)}$.

From the field equation for $\Phi$
\begin{equation}
\hat D_a \hat D_a \Phi - \Phi \lambda = 0,
\end{equation}
we can deduce
\begin{equation}
\lambda = \Phi^\dag \hat D_a \hat D_a \Phi = - (\hat D_a
\Phi)^\dag \hat D_a \Phi,
\end{equation}
and the field equation becomes
\begin{equation}
\hat D_a \hat D_a \Phi + \Phi (\hat D_a \Phi)^\dag \hat D_a \Phi
= 0.
\end{equation}

\section{BPS Solitons}

As in the (non-)commutative case, the $\cp$ model on the fuzzy
sphere has the Bogomolny bound. Let's consider the inequality
\begin{equation}
\half \Tr \left\{ (\hat D_i \Phi \pm i \epsilon_{ij} \hat D_j
\Phi)^\dag (\hat D_i \Phi \pm i \epsilon_{ik} \hat D_k \Phi)
\right\} + \Tr \left\{ (\hat D_3 \Phi)^\dag (\hat D_3 \Phi)
\right\} \ge 0,
\end{equation}
where $i,\,j=1,2$. Expanding this inequality we obtain
\begin{equation}
S = \Tr \left\{ (\hat D_i \Phi)^\dag (\hat D_i \Phi) + (\hat D_3
\Phi)^\dag (\hat D_3 \Phi) \right\} \ge \mp i \epsilon_{ij} \Tr
\left\{ (\hat D_i \Phi)^\dag (\hat D_j \Phi) \right\} \equiv \pm
Q,
\end{equation}
where the $U(1)$ gauge invariant ``topological charge'' is
\begin{equation}\la{Q}
Q = - i \epsilon_{ij} \Tr \left\{ (\hat D_i \Phi)^\dag (\hat D_j
\Phi) \right\} =  \Tr {\hat F}_{12}.
\end{equation}
The Bogomolny bound of the (Euclidean) action is saturated by the
configuration which satisfies the (anti-)self-dual equations
\begin{equation}\la{BPS1}
\hat D_i \Phi \pm i \epsilon_{ij} \hat D_j \Phi = 0, \qquad
\hbox{or} \qquad \hat D_\pm \Phi = 0,
\end{equation}
and
\begin{equation}\la{D3=0}
\hat D_3 \Phi = 0.
\end{equation}
The gauge covariant condition \eq{D3=0} is equivalent to
\eq{K3=k}. Since the field $\Phi$ corresponds to a section of the
bundle \eq{coset}, \eq{D3=0} is a natural requirement. In this way
we can naturally project the theory onto the fuzzy sphere by
specifying the topological sector of $\Phi$ although the field
$\Phi$ is defined on $\s^3$ rather than $\s^2$.

In terms of $W$ variables in \eq{W}, the (anti-)self-dual equation
reduces to
\begin{equation}\la{BPS2}
\hat D_\pm \Phi = P \left( \hat \nabla_\pm W \right) \left( W^\dag
W \right)^{-1/2} = 0,
\end{equation}
and the topological charge can be rewritten as
\begin{equation}\la{QW}
Q = \half \Tr \left\{ \frac1{W^\dag W} \left( \hat \nabla_+ W^\dag
P \hat \nabla_-W - \hat \nabla_- W^\dag P \hat \nabla_+ W\right)
\right\}.
\end{equation}
Eq.\eq{BPS2} is equivalent to $\hat \nabla_\pm W =W c(\hat r)$ for
an arbitrary complex function $c(\hat r)$. To find the
(anti-)self-dual configurations, the scale and the gauge
symmetries can be used to put $c(\hat r)=0$, ending with a pure
``(anti-)holomorphic equation''
\begin{equation}\la{BPS3}
\hat \nabla_\pm W = i[\hat K_\pm, W] = 0.
\end{equation}

Using the definitions of $\hat K_\pm$ operators in \eq{K+-}, one
can easily find the general BPS solutions satisfying \eq{D3=0} and
\eq{BPS3} as \footnote{The following formula may be useful to find
explicit solutions: $a f(a^\dag a)=f(a^\dag a+1)a,\;a^\dag
f(a^\dag a)=f(a^\dag a-1)a^\dag$ for operators $a$ and $a^\dag$
satisfying \eq{bi-harm} and a non-singular function $f(a^\dag
a).$}
\begin{eqnarray}\la{BPSsol}
W^+_k &=& \sum_{l=0}^{k} c^+_{kl} (a_1^\dag)^{l}
(a_2^\dag)^{k-l}\sqrt{\prod_{n=0}^{l-1}(N-n-a_1^\dag
a_1)(N-k+n-a_2^\dag a_2)}, \\
\la{antiBPSsol} W^-_k &=& \sum_{l=0}^{k} c^-_{kl}
\sqrt{\prod_{n=0}^{l-1}(N-n-a_1^\dag a_1)(N-k+n-a_2^\dag a_2)} \;
(a_1)^{l} (a_2)^{k-l},
\end{eqnarray}
where $k\leq N$. These solutions are eigenstates of $\hat K_3$
with the eigenvalue $\pm k$ and possess $2k(n+1)+2n$ real
parameters $c^{\pm}_{kl}$ to specify the BPS solutions.

Let's take the commutative limit, $\a \to 0$ (or $N \rightarrow
\infty$), where $a_1$ and $a_2$ become usual complex variables in
$\c^2$. And introduce the stereographic projection of the
(commutative) sphere to the complex plane given by
\begin{equation}\label{stereo}
  z=R\frac{r_1+ir_2}{R-r_3}=R\frac{a_1^\dag}{a_2^\dag},\qquad
  \bz=R\frac{r_1-ir_2}{R-r_3}=R\frac{a_1}{a_2},
\end{equation}
where the relations \eq{rL} and \eq{Ls} are used. Then the
solutions \eq{BPSsol} and \eq{antiBPSsol} can be rearranged into
the standard form \ct{ddl} in the commutative limit up to a scale
factor which can be scaled away using the scale symmetry ($W \to W
\Delta)$:
\begin{equation}\label{clsol}
W^+_k = a_{+k}^0\prod_{l=1}^{k} (z-a_{+k}^l), \qquad W^-_k =
a_{-k}^0\prod_{l=1}^{k} (\bz-a_{-k}^l).
\end{equation}
Thus it is clear that the parameters $c_{kl}^{\pm}$ in \eq{BPSsol}
can be interpreted as the moduli of $k$ BPS solitons on the fuzzy
sphere and the dimension of moduli space of $k$ solitons is
exactly the same as that of the commutative sphere or the
(noncommutative) plane.

The vacuum moduli space has $2n$ parameters for $\cp$ space but,
in commutative and large radius limit, they have infinite inertia
due to the volume factor. For fuzzy sphere, since the volume is
finite and there are only finite number of states, it is
interesting to study the moduli space dynamics of solitons,
including these vacuum moduli.

Now let's calculate the topological charge $Q$ defined by \eq{QW}
for the BPS solutions \eq{BPSsol}. For the configurations
satisfying \eq{BPS3}, we first note that Eq.\eq{QW} can be
rewritten as
\begin{equation}\la{Qs}
Q_s =\half \Tr \left\{\hat \nabla_+ \left( \frac1{W^\dag W} W^\dag
\hat \nabla_- W\right) - \frac1{W^\dag W} W^\dag \left(\hat
\nabla_+ \hat \nabla_- -\hat \nabla_- \hat \nabla_+ \right)W
\right\}
\end{equation}
for the soliton \eq{BPSsol} satisfying $\hat \nabla_+ W = 0$, and
\begin{equation}\la{Qas}
Q_{as} = \half \Tr \left\{-\hat \nabla_- \left( \frac1{W^\dag W}
W^\dag \hat \nabla_+ W\right) +\frac1{W^\dag W} W^\dag \left(\hat
\nabla_+ \hat \nabla_- -\hat \nabla_- \hat \nabla_+ \right)W
\right\}
\end{equation}
for the anti-soliton \eq{antiBPSsol} satisfying $\hat \nabla_- W =
0$. On the fuzzy sphere, the traces of total derivative in
\eq{Qs} and \eq{Qas} are always zero because $\CH_N$ is of finite
dimension, while, on the (noncommutative) plane, the topological
charge is coming from them \ct{LLY}. On the other hand, the
second terms in \eq{Qs} and \eq{Qas} on the fuzzy sphere become
$\mp 2i\hat \nabla_3$, which gives the topological charge, while,
on the (noncommutative) plane, they vanish. Thus we have shown
that the topological charge \eq{Qs} or \eq{Qas} is an integer
number given by an eigenvalue of $\hat K_3$;
\begin{equation}\label{topcharge}
Q_s= k, \qquad Q_{as}= -k.
\end{equation}

In section 6 we will speculate on a topological issue related to
the charge $Q$ of BPS solitons on fuzzy sphere \ct{charge}.

\section{Atiyah-Singer Index on Fuzzy Sphere}

In this section we will calculate the zero modes of Dirac operator
on fuzzy sphere under the BPS background defined by \eq {BPS1} and
\eq{D3=0}. We will show that the Atiyah-Singer index, that is the
number of the zero modes of Dirac operator, is exactly given by
the topological charge $|Q|$. Thus this result presumably implies
that the Atiyah-Singer index theorem is still valid on fuzzy
sphere. (For discussions about the Atiyah-Singer index on
commutative sphere in the case of gauge theory, see \ct{NS}.)

We will take a spinor field $\Psi(\hat r)$ as follows
\begin{equation}\label{spinor}
  \Psi(\hat r) = \Psi^+(\hat r) b +\Psi^-(\hat r) b^\dag,
\end{equation}
where $\Psi^{\pm}(\hat r)$ are bosonic fields defined on the fuzzy
sphere and $b, b^\dag$ are Grassmannian operators satisfying the
following anti-commutation relations
\begin{equation}\label{bb}
 \{b,b\}=\{b^\dag, b^\dag\}=0, \qquad \{b,b^\dag\}=1.
\end{equation}
As shown in \ct{GKP} and \ct{Pres}, using these anticommuting
operators $b$ and $b^\dag$, the Clifford algebra appropriate for
the fuzzy sphere can be constructed:
\begin{equation}\label{Clifford}
  \Gamma_+ = b \{b, \bullet\}, \quad
  \Gamma_- = b^\dag \{b^\dag, \bullet\}, \quad
  \Gamma_3 = b \{b^\dag, \bullet\}
  - b^\dag \{b, \bullet\},
\end{equation}
where the Gamma matrices $\Gamma_{\pm}, \Gamma_3$ act on a
Grassmannian operator $\zeta= b\; \mbox{or}\; b^\dag$, e.g.
$\Gamma_+\zeta = b \{b, \zeta\}$, etc. Then it is easy to check
the Clifford algebra
\begin{equation}
  \Gamma_{\pm}^2=0, \qquad \Gamma_3^2=1,
\end{equation}
and
\begin{equation}\label{Calg}
\Gamma_+\Gamma_- + \Gamma_-\Gamma_+ =1, \qquad
\Gamma_{\pm}\Gamma_3 + \Gamma_3\Gamma_{\pm} =0, \qquad
\Gamma_+\Gamma_- - \Gamma_-\Gamma_+ = \Gamma_3.
\end{equation}
Thanks to the last relation in \eq{Calg}, we can regard $\Gamma_3$
as the chirality operator on the fuzzy sphere. So $\Psi^{+}$ and
$\Psi^-$ in \eq{spinor} are the components with the chirality
$+1$ and $-1$ respectively.

Now we will consider massless fermions interacting with the $\cp$
fields satisfying \eq{BPS1} and \eq{D3=0}. The fermions $\Psi$
are vectors in $\c^{n+1}$ like $\Phi$ constrained by \ct{ddl}
\begin{equation}\label{PhiPsi=0}
\Phi^\dag \Psi = {\bar \Psi} \Phi=0,
\end{equation}
where ${\bar \Psi}=\Gamma_3 \Psi^\dag=(\Psi^-)^\dag b -
(\Psi^+)^\dag b^\dag$. For the purpose of manifestly $SU(2)$
invariant action as in section 3, we will first consider the Dirac
action on $\s^3$ (constructed by the noncommutative $\c^2$
generated by \eq{bi-harm} with the constraint $\xi^\dag \xi =N$)
and later project the theory onto fuzzy sphere. The covariant
derivative about a spinor $\Psi$ coupled to the gauge field $\hat
A_a$ given by \eq{A} is defined as follows
\begin{equation}\label{coder}
\CD\Psi=\Gamma^a \hat D_a \Psi= \Gamma^a(\hat \nabla_a\Psi - i\Psi
\hat A_a).
\end{equation}
The relevant Dirac action for our problem turns out to be
\footnote{Here we are implicitly assuming the trace over the
fermionic Fock space generated by $b$ and $b^\dag$,
$\{|\nu\rangle, \;\nu=0,1\}$.}
\begin{equation}\label{Dirac-action}
S_D=\Tr \left[ {\bar \Psi} i\CD \Psi + {\bar \lambda} \Phi^\dag
\Psi + {\bar \Psi}\Phi \lambda \right],
\end{equation}
where ${\bar \lambda}$ and $\lambda$ are Lagrange multipliers to
incorporate the constraints \eq{PhiPsi=0}. The above Dirac action
has a global $U(n+1)$ symmetry and a local $U(1)$ symmetry
\begin{equation} \label{GT3}
\Psi(\hat r) \to \Psi(\hat r) g(\hat r),
\end{equation}
together with the transformation \eq{GT1}.

From the equation of motion for $\Psi$,
\begin{equation}
  i\CD \Psi + \Phi \lambda=0,
\end{equation}
we can deduce
\begin{equation}
  \lambda=-i \Phi^\dag \CD \Psi,
\end{equation}
and the resulting Dirac equation becomes
\begin{equation}\label{Diraceq}
(1-\Phi\Phi^\dag)\CD \Psi =0 \quad \mbox{or}\quad P\CD \Psi =0,
\end{equation}
where $P$ is the projection operator defined by \eq{P}.

The constraints in \eq{PhiPsi=0} can be solved by introducing a
spinor, $\eta(\hat r)=\eta^+(\hat r) b +\eta^-(\hat r) b^\dag$,
of the form
\begin{equation}\label{nucon-spinor}
\Psi=P\eta,
\end{equation}
where the spinor $\eta$ is an unconstrained vector in $\c^{n+1}$.
In terms of $\eta$ variables, the action \eq{Dirac-action} can be
rewritten as
\begin{equation}\label{Diraction}
  S_D=\Tr \left\{ {\bar \eta} P i\CD (P\eta) \right\}.
\end{equation}
The equation of motion in terms of $\eta$ becomes
\begin{equation}\label{eom}
P \CD (P\eta)=0.
\end{equation}
It is obvious that this action has the $U(1)$ gauge symmetry as
well as an additional symmetry given by the shift of fermions,
\begin{equation}\label{shift}
  \eta \to \eta+W \chi,
\end{equation}
where $\chi=\chi^+ b + \chi^-b^\dag$ is an arbitrary spinor. This
shift symmetry is a fermionic partner of the scale symmetry in
the bosonic action \eq{Waction}.

The fields $\Psi^{\pm}$ can be expanded on the noncommutative
$\c^2$ similarly to \eq{Phi} \ct{GKP}
\begin{equation}\label{Psi}
\Psi^{\pm}=\sum
\Psi_{m_1m_2n_1n_2}^{\pm}{a_1^\dag}{}^{m_1}{a_2^\dag}{}^{m_2}
a_1^{n_1}a_2^{n_2}.
\end{equation}
The chiral fields $\Psi^{\pm}$ can also be classified according to
the $U(1)$ gauge transformation \eq{U1} and the set of the fields
with definite $U(1)$ charge $k$ will be denoted as
$\Psi^{\pm}_k$. Then the chiral fields $\Psi^{\pm}_k$ transform
under the $U(1)$ gauge transformation \eq{U1} as
\begin{equation}\label{U1tfs}
  \Psi^{\pm}_k \;\;\to \;\;\Psi^{\pm}_k e^{-ik\psi}.
\end{equation}
Now we will take gauge covariant projection onto the fuzzy sphere:
\begin{equation}\la{D3s=0}
\hat D_3 \Psi_k =0,
\end{equation}
which is equivalent to
\begin{equation}\label{K3s=ks}
[\hat K_3, \Psi_k]=k\Psi_k.
\end{equation}
In the projected subspace \eq{D3s=0}, the action is invariant
with respect to the chiral transformation \ct{Pres}
\begin{equation}\label{chiraltr}
  \Psi_k(\hat r) \to e^{i\alpha \Gamma_3}\Psi_k(\hat r), \qquad
  {\bar \Psi}_k(\hat r) \to {\bar \Psi}_k(\hat r)e^{i\alpha \Gamma_3}
\end{equation}
which is the result of the second Clifford algebra in \eq{Calg}.

We will solve the equation of motion \eq{eom} for $\eta_k$
obeying $\hat D_3 \eta_k =0$ under the soliton background
\eq{BPSsol} satisfying $\hat \nabla_+ W = 0$. (The analysis with
the anti-soliton background \eq{antiBPSsol} ($\hat \nabla_- W =
0$) will be similar to the soliton case.) The equation \eq{eom}
under this background has the following component form \bea
\label{Diraceq+} &&P\left([\hat K_+,\eta^+_k]-\eta^+_k\sqrt{W^\dag
W}[\hat K_+, \frac{1}{\sqrt{W^\dag
W}}]\right)=0,\\
\la{Diraceq-} && P\left([\hat K_-,\eta^-_k]-[\hat K_-,
W]\frac{1}{W^\dag W}W^\dag \eta^-_k +\eta^-_k[\hat K_-,
\frac{1}{\sqrt{W^\dag W}}]\sqrt{W^\dag W}\right)=0, \eea where
$PW=0$ and $[\hat K_-, W^\dag]=0$ are used.

It is easy to see that the solution of the positive chirality
$\eta^+_k$ up to the shift symmetry \eq{shift} is given by
\begin{equation}\label{sol+}
\eta^+_k = \zeta^+_k (a_1^\dag, a_2^\dag)\frac{1}{\sqrt{W^\dag
W}},
\end{equation}
where $\zeta^+_k$ is a positive chirality spinor satisfying $[\hat
K_+, \zeta^+_k]=0$ and $[\hat K_3, \zeta^+_k]=k\zeta^+_k$:
\begin{equation}\label{zeta+}
\zeta^+_k=\zeta^+_{kl} (a_1^\dag)^{l}
(a_2^\dag)^{k-l}\sqrt{\prod_{n=0}^{l-1}(N-n-a_1^\dag
a_1)(N-k+n-a_2^\dag a_2)}, \qquad l=0,\cdots,k.
\end{equation}
And it is also easy to check that the solution for $\eta^-_k$ with
the negative chirality is given by
\begin{equation}\label{sol-}
\eta^-_k = W \zeta^-_k,
\end{equation}
where $\zeta^-_k$ is an arbitrary negative chirality spinor
satisfying $[\hat K_3, \zeta^-_k]=k\zeta^-_k$. However, the
solution \eq{sol-} can be gauged away using the shift symmetry
\eq{shift} or $\Psi^-_k=P\eta^-_k=0$, so there are no zero modes
(up to the gauge and the shift symmetry) for negative chirality
spinors under the soliton background.

Thanks to the finite volume of fuzzy sphere, the normalizability
of zero modes \eq{zeta+} is automatically guaranteed. Note that
$k+1$ zero modes in \eq{zeta+} are not independent each other. The
number of independent zero modes is so $k$, not $k+1$. The reason
is following. If we take a linear combination of these zero modes
for $\eta^+_k$ to be proportional to $W$ (there is only one such
combination since we should take $\zeta^+_{kl}=c_{kl}^+{\chi}^+_k$
for all $l$), it is a trivial solution since $\Psi^+_k= P \eta^+_k
=PW {\chi}^+_k/\sqrt{W^\dag W}=0$. We have shown in section 4 that
this integer number is exactly the topological charge of
background solitons. Thus we arrived at the Atiyah-Singer index
theorem on fuzzy sphere.

For a spin complex, the Atiyah-Singer index theorem \ct{Nakahara}
states that
\begin{equation}\la{AS-index}
\mbox{Index}\,{\hat D} = \mbox{dim}\;\mbox{ker}{\hat
D}-\mbox{dim}\;\mbox{ker}{\hat D}^\dag= n_+- n_-,
\end{equation}
where ${\hat D}$ is the Dirac operator for the spin complex and
$n_{\pm}$ is the number of normalizable zero modes of the Dirac
operator of chirality $\pm 1$. For a monopole bundle $P(\s^2,
U(1))$ which corresponds to our case \eq{coset}, the expression
\eq{AS-index} reads
\begin{equation}\label{monopole-bundle}
  n_+ - n_-=\frac{1}{2\pi}\int_{\s^2} F \in {\bf Z},
\end{equation}
where $F=dA$ is the field strength of a (monopole) gauge field and
the integer quantization is coming from the homotopy
$\pi_1(U(1))={\bf Z}$ \ct{Nakahara}. In this paper, we proved the
noncommutative version of \eq{monopole-bundle}:
\begin{equation}\label{fsAS-index}
  n_+ =\Tr {\hat F}_{12}= Q,
\end{equation}
where $n_-=0$ for the soliton background and the $U(1)$ field
strength $\hat F_{12}$ is given by \eq{F}.

\section{Discussion}

In this paper we showed that the $\cp$ model on fuzzy sphere
enjoys all attractive properties in commutative space. The BPS
equations support (anti-)self-dual soliton solutions and the
dimension of moduli space of BPS solitons is exactly the same as
the commutative case. Moreover, the number of normalizable zero
modes in the presence of the soliton backgrounds is exactly given
by the topological charge of the solitons, thus the Atiyah-Singer
index theorem remains valid even for fuzzy sphere. This seems to
be in contrast to the recent results in \cite{Noninteger}
claiming that the $U(1)$ monopole charge is not integer for the
fuzzy sphere at finite cut-off $N$. A further investigation on
the difference between their case and ours should be interesting.

On fuzzy sphere, only finite number of states can be defined
\ct{Madore}. For example, for the fuzzy sphere with a cut off spin
$N$, only $N+1$ states are distinguishable. So it seems to be
reasonable that one cannot put too many solitons on fuzzy sphere.
Moreover the generators ${\hat K}_{\pm}$ in the Holstein-Primakoff
realization contain the square-root factors such as $\sqrt{N-
a_1^\dag a_1}$ and $\sqrt{N- a_2^\dag a_2}$. Thus in order to
preserve the theory to be unitary, we should have an upper bound
on the occupation numbers, i.e. $a_1^\dag a_1, a_2^\dag a_2 \leq
N$. Then this also put an upper bound on the topological charge
$|Q|=k$ since it is related to the ${\hat K}_3$ eigenvalue of a
BPS solution $\Phi_k$. As this speculation implies, more careful
study is certainly required to understand the topological nature
of fuzzy sphere. The topological properties of fuzzy sphere have
been studied in \ct{charge} based on the boson realization of
$SU(2)$ algebra, Schwinger vs. Holstein-Primakoff.

The outstanding properties of (supersymmetric) $\cp$ model with
quarks are asymptotic freedom, confinement of quarks, and
spontaneous chiral symmetry breaking, which are very similar to
QCD$_4$ \ct{ddl,Witten}. Since the $\cp$ model can be $1/n$
expanded, all these properties can be explored based on the $1/n$
expansion. A natural way to couple quarks to $\cp$ fields is to
introduce the supersymmetric $\cp$ model. The $\cp$ model on
noncommutative spaces presented here and in \ct{LLY} can be
generalized to the supersymmetric model. It is interesting to
study low-energy dynamics of quarks in the context of
supersymmetric $\cp$ model on noncommutative spaces (plane and
sphere) since it is more similar to QCD$_4$ due to the non-Abelian
nature from noncommutative space. As indicated in section 5, axial
anomaly and related $U(1)_A$ problem can also be studied along
this line.

The (noncommutative) $\cp$ model can be understood as a formal
limit of (noncommutative) Maxwell-Higgs theory,
\[
S = \Tr \left[ \frac1{4g^2} \hat F_{ab}\hat F^{ab} + (\hat D_a
\Phi)^\dag (\hat D_a \Phi) + \half \lambda (\Phi^\dag \Phi - 1)^2
\right],
\]
namely, $\lambda\rightarrow\infty$ and $g^2\rightarrow\infty$.
Thus this theory defined on fuzzy sphere may also enjoy the same
properties as the $\cp$ model even though the dynamics of gauge
fields is considered. This model, more generally,
Maxwell-Chern-Simons-Higgs theory on fuzzy sphere, can be of
interest itself since it is related to the world-volume theory of
the spherical D2-brane formed by the bound state of $N$ D0-branes
\ct{ars}.

It would be desirable to extend the analysis in this paper to
four-dimensional case, especially instantons on fuzzy $\s^4$. On
the noncommutative $\r^4$, it was shown that the noncommutative
instantons are well defined where small instanton singularities
are resolved \ct{ns} and the topological charge of instantons is
always an integer \ct{KLY}. On fuzzy $\s^4$, the topological
properties of instanton solutions may appear with more elegant
structures.

\section*{Acknowledgments}
We are grateful to Pei-Ming Ho and Miao Li for helpful
discussions. FLL likes to thank Bin Chen for discussions. HSY also
thanks Bum-Hoon Lee and Kimyeong Lee for discussions. Two of us
are supported by NSC (CTC: NCS89-2112-M-009-006 and HSY:
NCS89-2811-M-002-0095). CMC is supported by the CosPA project of
the Ministry of Education, Taiwan. FLL is supported by BK-21
Initiative in Physics (SNU-Project 2). We also acknowledge NCTS
as well as CTP at Taida for partial support.

\appendix

\section*{Appendix}

\section{Fuzzy Spherical Harmonics}

First we will briefly review quantum mechanics on the addition of
angular momentum in order to fix the notations and to illustrate
how to generalize it to fuzzy spherical harmonics. We find the
most useful reference on this is \cite{J}.

The space Fun($\s^2$) of functions on $\s^2$ is spanned by
spherical harmonics $Y^J_m \in {\mbox Fun}(\s^2)$ where $J$ runs
through all integer spins. A product of any two spherical
harmonics is again a function on $\s^2$ and hence it can be
written as a linear combination of spherical harmonics,
\begin{equation}\label{prodYY}
  Y^I_lY^J_m = \sum_{K,n} \sqrt{\frac{(2I+1)(2J+1)}{4\pi(2K+1)}}
  c_{IJK}C^{IJK}_{lmn}Y_n^K
\end{equation}
with $C^{IJK}_{lmn}$ denoting the Clebsch-Gordan coefficients of
  $su(2)$. The explicit form of structure constants
  $c_{IJK}=C^{IJK}_{000}$
  is given by \cite{J}
\begin{equation}\label{cIJK}
  c_{IJK}=\Biggl\{
\begin{array}{cc}
  0, & \mbox{if} \; I+J+K=2g+1, \\
  \frac{(-1)^{g-K}\sqrt{2K+1} g!}{(g-I)!(g-J)!(g-K)!}
  \Bigl[\frac{(2g-2I)!(2g-2J)!(2g-2K)!}{(2g+1)!} \Bigr]^{1\over 2}, &
  \mbox{if} \; I+J+K=2g,
\end{array}
\end{equation}
where $g$ is a positive integer.

The spherical harmonics $Y_m^J$ form multiplets with respect to
the $su(2)$ action on Fun($\s^2$). More generally, the direct
product of two irreducible tensors ${\mathcal{M}}^I_l$ and
${\mathcal{N}}^J_m$ may be decomposed into irreducible tensors.
The coefficients of this decomposition are just the Clebsch-Gordan
coefficients:
\begin{equation}\label{MN}
  {\mathcal{M}}^I_l {\mathcal{N}}^J_m=\sum_{K,n}
  C^{IJK}_{lmn}\{{\mathcal{M}}\otimes {\mathcal{N}}\}_n^K.
\end{equation}
The inverse relation is
\begin{equation}\label{inverseMN}
  \{{\mathcal{M}}\otimes {\mathcal{N}}\}_n^K=\sum_{l,m}
  C^{IJK}_{lmn}{\mathcal{M}}^I_l {\mathcal{N}}^J_m.
\end{equation}

The fuzzy sphere is constructed replacing the algebra of function
on $\s^2$, Fun($\s^2$), by the noncommutative algebra taken in an
irreducible representation of $SU(2)$. This is full matrix algebra
Mat($N$+1) which is generated by the fuzzy spherical harmonics
${\hat Y}_m^J$ with $-J\leq m \leq J,\; J=0,1, \cdots, N$, a
complete basis of the space Mat($N$+1). The explicit form of
${\hat Y}_m^J$ in $(N+1)$-dimensional representation is given by
\cite{J}
\begin{equation}\label{matY}
  [{\hat Y}_m^J]_{s's}=\sqrt{\frac{2J+1}{N+1}}C^{NJN}_{sms'},
\end{equation}
where $s,\,s'=-{N\over 2},\cdots,0,\cdots,{N\over 2}$. The
operators ${\hat Y}_m^J$ transform under $su(2)$ according to the
representation $D^J$, so they are irreducible tensors of rank $J$.
Thus an arbitrary matrix ${\hat A} \in \mbox{Mat}(N+1)$ may be
written in the form
\begin{equation}\label{opA}
  {\hat A}=\sum_{J=0}^N \sum_{m=-J}^J A_{Jm}{\hat Y}_m^J
\end{equation}
where the expansion coefficients $A_{Jm}$ are given by
\begin{equation}\label{AJm}
  A_{Jm} =\Tr ({\hat Y}^{\dag J}_m {\hat A}).
\end{equation}
The derivative of ${\hat A} \in \mbox{Mat}(N+1)$ is defined by the
adjoint action of ${\hat S}_a \in su(2)$
\begin{equation}\label{der}
  {\hat \nabla}_a{\hat A}=i[{\hat S}_a, {\hat A}]=
  i\sum_{J=0}^N \sum_{m=-J}^J A_{Jm}[{\hat S}_a,{\hat Y}_m^J].
\end{equation}
The commutators in \eq{der} can be calculated by the following
commutation relations \cite{J}
\begin{equation}\label{commutator}
 [{\hat S}_\mu, {\hat Y}_{m}^J]=\sqrt{J(J+1)}C_{m \mu (m+\mu)}^{J1J}
 {\hat Y}_{m+\mu}^J,\quad(\mu=\pm,0,\;{\hat S}_0\equiv {\hat S}_3).
\end{equation}
The Laplacian on the fuzzy sphere is given by
\begin{equation}\label{lap}
  {\hat \nabla}_a {\hat \nabla}_a{\hat A}=
  -\sum_{J=0}^N \sum_{m=-J}^J A_{Jm}[{\hat S}_a,[{\hat S}_a,{\hat
  Y}_m^J]]= -\sum_{J=0}^N \sum_{m=-J}^J J(J+1)A_{Jm}{\hat Y}_m^J.
\end{equation}

The product of any two such matrices can be expressed as a linear
combination of matrices ${\hat Y}_n^K$ \cite{J},
\begin{equation}\label{opYY}
  {\hat Y}^I_l{\hat Y}^J_m =
  \sum_{K=0}^N(-)^{N+K}\sqrt{(2I+1)(2J+1)} \Bigl\{\begin{array}{ccc}
    I & J & K \\
    {N\over 2} & {N\over 2} & {N\over 2} \
  \end{array}
\Bigr\}C^{IJK}_{lmn}{\hat Y}_n^K
\end{equation}
where $\{\begin{array}{ccc}
    \cdot & \cdot & \cdot \\
    \cdot & \cdot & \cdot \
  \end{array}\}$ denotes the recoupling coefficients
  ($6J$-symbols) of $su(2)$.
The inverse relation is
  \begin{equation}\label{opY}
  {\hat Y}^K_n=
  \sum_{IlJm}(-)^{N+I+J}\sqrt{(2I+1)(2J+1)} \Bigl\{\begin{array}{ccc}
    I & J & K \\
    {N\over 2} & {N\over 2} & {N\over 2} \
  \end{array}
\Bigr\}C^{IJK}_{lmn}{\hat Y}_l^I{\hat Y}_m^J.
\end{equation}
More generally, the product module $V^I \bigotimes V^J$ can be
expanded into the irreducible module $V^K$:
\begin{equation}\label{decomp}
  V^I \bigotimes V^J = \bigoplus_{K=0}^{N}V^K.
\end{equation}

Note that the $SU(2)$ operators ${\hat K}$ in \eq{K3} and \eq{K+-}
are a sum of two spin operators ${\hat S}_{1}$ and ${\hat S}_{2}$:
\begin{equation}\label{K}
  {\hat K}={\hat S}_{1}+{\hat S}_{2},
\end{equation}
where
\begin{equation}\label{SS}
  [{\hat S}_{1a},{\hat S}_{1b}]=i\epsilon_{abc}{\hat S}_{1c},
  \quad [{\hat S}_{2a},{\hat S}_{2b}]=i\epsilon_{abc}{\hat
  S}_{2c}, \quad [{\hat S}_{1a},{\hat S}_{2b}]=0.
\end{equation}
The $(N+1)$-dimensional unitary representations of ${\hat S}_{1}$
and ${\hat S}_{2}$ in the Holstein-Primakoff realization
\cite{HP}, denoted as ${\cal H}_N^1$ and ${\cal H}_N^2$,
respectively, can be given by the following orthonormal basis
\begin{eqnarray}\label{HH}
{\cal H}_N^1&=&\{|n\rangle_1 =|\frac{N}2, n-\frac{N}2\rangle_1 =
(a_1^\dag)^{n}|0\rangle_1,\; n=0,1,\cdots, N\},\nonumber\\
{\cal H}_N^2&=&\{|m\rangle_2 =|\frac{N}2, m-\frac{N}2\rangle_2=
(a_2^\dag)^{m}|0\rangle_2,\; m=0,1,\cdots, N\},
\end{eqnarray}
where $|l, m\rangle_{1,2}$ is a spherical harmonics for each spin
operator and $|0\rangle_{1,2}$ is the vacuum defined by
$a_1|0\rangle_1=a_2|0\rangle_2=0$. Then the basis \eq{basis} is a
tensor product of ${\cal H}_N^1$ and ${\cal H}_N^2$ and can be
expanded in the basis of total spin operator, that is,
\begin{equation}\label{decH}
{\cal H}_N^1\bigotimes{\cal H}_N^2=\bigoplus_{J=0}^{N}{\cal
H}^{(J)}
\end{equation}
and
\begin{equation}\label{H}
\bigoplus_J{\cal H}^{(J)}=\{|J, 0\rangle, \; J=0,1,\cdots, N\},
\end{equation}
where the spherical harmonics $|J, 0\rangle$ is an irreducible
basis of the total spin operator ${\hat K}$:
\begin{equation}\label{totalbasis}
  \hat K^2|J, 0\rangle=J(J+1)|J, 0\rangle, \;\;\;
  \hat K_3|J, 0\rangle=0.
\end{equation}
The second condition in \eq{totalbasis} is coming from the usual
rule of the addition of angular momentum, $m=m_1+m_2$, which is
zero for the basis \eq{basis}, where ${\hat
S}_{13}|N/2,m_1\rangle_1=m_1|N/2,m_1\rangle_1,\;{\hat
S}_{23}|N/2,m_2\rangle_2=m_2|N/2,m_2\rangle_2,$ and ${\hat
K}_{3}|J,m\rangle=m|J,m\rangle$. Thus the states in \eq{H} can
serve as $(N+1)$-dimensional basis of ${\hat K}$.

Let ${\hat Y}_{1l}^I,\,{\hat Y}_{2m}^J$, and ${\hat Y}_n^K$ be the
fuzzy spherical harmonics for ${\hat S}_1,\,{\hat S}_2$, and
${\hat K}$, respectively. They are complete and irreducible basis
of the space Mat$(N+1)$ whose matrix elements can be represented
in the corresponding basis ${\cal H}_N^1,\,{\cal H}_N^2$ and
${\cal H}^{(J)}$ and are given by \eq{matY}. Furthermore they
satisfy the Clebsch-Gordan decomposition \eq{opYY} and \eq{opY}.
Thus, using these relations, the Casimir operator of ${\hat K}$ as
in \eq{lap} can be calculated based on these basis:
\begin{eqnarray}\label{Casimir}
 [{\hat K}_a,[{\hat K}_a,{\hat Y}^K_n]]&=&K(K+1){\hat Y}^K_n
 \nonumber\\
&=&\sum_{IlJm}(-)^{N+I+J}\sqrt{(2I+1)(2J+1)}
\Bigl\{\begin{array}{ccc}
    I & J & K \\
    {N\over 2} & {N\over 2} & {N\over 2} \
  \end{array}
\Bigr\}C^{IJK}_{lmn}\times \nonumber\\
&&\Bigl(\bigl(I(I+1)+J(J+1)\bigr){\hat Y}_{1l}^I{\hat
Y}_{2m}^J+2[{\hat S}_{1a},{\hat Y}_{1l}^I][{\hat S}_{2a},{\hat
Y}_{2m}^J]\Bigr),
\end{eqnarray}
where $[{\hat S}_{1},{\hat Y}_{2m}^J]=[{\hat S}_{2},{\hat
Y}_{1l}^I]=0$ are used. The commutators in \eq{Casimir} can be
calculated by using \eq{commutator}.

\newpage


\nc{\np}[3]{Nucl. Phys. {\bf B#1} (#2) #3}

\nc{\pl}[3]{Phys. Lett. {\bf B#1} (#2) #3}

\nc{\prl}[3]{Phys. Rev. Lett.{\bf #1} (#2) #3}

\nc{\prd}[3]{Phys. Rev. {\bf D#1} (#2) #3}

\nc{\ap}[3]{Ann. Phys. {\bf #1} (#2) #3}

\nc{\prep}[3]{Phys. Rep. {\bf #1} (#2) #3}

\nc{\rmp}[3]{Rev. Mod. Phys. {\bf #1} (#2) #3}

\nc{\cmp}[3]{Comm. Math. Phys. {\bf #1} (#2) #3}

\nc{\mpl}[3]{Mod. Phys. Lett. {\bf #1} (#2) #3}

\nc{\cqg}[3]{Class. Quant. Grav. {\bf #1} (#2) #3}

\nc{\jhep}[3]{J. High Energy Phys. {\bf #1} (#2) #3}

\nc{\hep}[1]{{\tt hep-th/{#1}}}



\begin{thebibliography}{99}

\bibitem{cds} A. Connes, M. R. Douglas, and A. Schwarz,
\jhep {02}{1998}{003} ; M. R. Douglas and C. Hull, {\it ibid.}
{\bf 02} (1998) 008.


\bibitem{ho} P.-M. Ho and Y.-S. Wu, \prd {58}{1998}{066003};
P.-M. Ho, {\it ibid.} {\bf B434} (1998) 41; A. Schwarz, \np {534}
{1998}{720} ; D. Brace, B. Morariu, and B. Zumino, {\it ibid.}
{\bf B545} (1999) 192; C. Hofman and E. Verlinde, {\it ibid.}
{\bf B547} (1999) 157; E. Kim, H. Kim, N. Kim, B.-H. Lee, C.-Y.
Lee, and H. S. Yang, \prd {62}{2000}{046001}.


\bibitem{aas} F. Ardalan, H. Arfaei, and M. M. Sheikh-Jabbari,
\jhep {02}{1999}{016}; C.-S. Chu and P.-M. Ho, \np {550}{1999}
{151}; V. Schomerus, \jhep {06}{1999}{030}.


\bibitem{sw} N. Seiberg and E. Witten, \jhep {09}{1999}{003}.



\bibitem{bell} J. Bellissard, K-theory of $C^*$-algebras in solid
    state physics, Lecture Notes in Physics 247 (1986) 99;
L. Susskind, The Quantum Hall Fluid and Non-Commutative Chern
Simons Theory, \hep{0101029}.



\bibitem{Madore} J. Madore, \cqg {9}{1992}{69}.



\bibitem{Hoppe} B. de Wit, J. Hoppe, and H. Nicolai, \np {305}{1988}{545};
J. Hoppe, Int. J. Mod. Phys. {\bf A4} (1989) 5235.



\bibitem{holi} A. Jevicki and S. Ramgoolam, \jhep {04}{1999}{032};
P.-M. Ho, S. Ramgoolam, and R. Tatar, \np {573}{2000}{364}; R.
Myers, \jhep {12}{1999}{022}; J. McGreevy, L. Susskind, and N.
Toumbas, \jhep {06}{2000}{008}; M. Li, \prd {63}{2001}{086002};
P.-M. Ho and M. Li, \np{596}{2001}{259}.



\bibitem{ars} C. Bachas, M. Douglas, and C. Schweigert, \jhep {05}{2000}{048};
A. Yu. Alekseev, A. Recknagel, and V. Schomerus, {\it ibid.} {\bf
09} (1999) 023; {\it ibid.} {\bf 05} (2000) 010; Y. Hikida, M.
Nozaki, and T. Takayanagi, \np {595}{2001}{319}; Y. Hikida, M.
Nozaki, and Y. Sugawara, Formation of Spherical D2-brane from
Multiple D0-branes, \hep{0101211}; K. Hashimoto and K. Krasnov,
\prd {64}{2001}{046007}.



\bibitem{WZW} S. Stanciu, \jhep {10}{2000}{015};
A. Yu. Alekseev and V. Schomerus, RR charges of D2-branes in the
WZW model, \hep{0007096}; J. M. Figueroa-O'Farrill and S.
Stanciu, \jhep {01}{2001}{006}.



\bibitem{iktw} Pei-Ming Ho, \jhep {12}{2000}{015};
S. Iso, Y. Kimura, K. Tanaka, and K. Wakatsuki, \np {604}{2001}
{121}.



\bibitem{K-theory} S. Fredenhagen and V. Schomerus,
\jhep {04}{2001}{007}.



\bibitem{ftfs} H. Grosse and J. Madore, \pl {283}{1992}{218};
H. Grosse, C. Klimcik, and P. Presnajder, Int. J. Theor. Phys.
{\bf 35} (1996) 231; \cmp {185}{1997}{155}; U. Carow-Watamura and
S. Watamura, {\it ibid.} {\bf 183} (1997) 365; Int. J. Mod. Phys.
{\bf A13} (1998) 3235; \cmp {212}{2000}{395}; C. Klimcik, {\it
ibid.} {\bf 199} (1998) 257; S. Baez, A. P. Balachandran, S.
Vaidya, and Y. Ydri, {\it ibid.} {\bf 208} (2000) 787; A. P.
Balachandran and S. Vaidya, Int. J. Mod. Phys. {\bf A16} (2001) 1.




\bibitem{GKP} H. Grosse, C. Klimcik,
and P. Presnajder, \cmp {178}{1996}{507}.



\bibitem{Pres} P. Presnajder, J. Math. Phys. {\bf 41} (2000) 2789.



\bibitem{HP} T. Holstein and H. Primakoff, Phys. Rev. {\bf 58}
(1940) 1098.


\bibitem{BL} B. Berg and M. L\"uscher, \cmp {69}{1979}{57}.




\bibitem{LLY} B.-H. Lee, K. Lee, and H. S. Yang,
\pl {498}{2001}{277}.



\bibitem{Jacob} N. Jacobson, Lie Algebras
(John Willy \& Sons, New York, 1962).



\bibitem{gpe} V. Golo and A. Perelomov, Phys. Lett. {\bf 79B} (1978) 112;
H. Eichenherr, \np {146}{1978}{215}.


\bibitem{ddl} A. D'Adda, P. Di Vecchia, and M. L\"uscher,
\np {146}{1978}{63}; {\it ibid.} {\bf B152} (1979) 145.


\bibitem{Witten} E. Witten, \np {149}{1979}{285}.


\bibitem{charge} C.-T. Chan, C.-M. Chen, and H. S. Yang,
Topological ${\bf Z}_{N+1}$ Charges on Fuzzy Sphere,
\hep{0106269}.


\bibitem{NS} N. K. Nielsen and B. Schroer, \np {127}{1977}{493}.


\bibitem{Nakahara} See, for example, M. Nakahara, {\it Geometry,
Topology and Physics} (Adam Hilger, Bristol and New York, 1990).



\bibitem{Noninteger} P. Valtancoli, \mpl {A16}{2001}{639};
H. Grosse and C. W. Rupp, A Remark on Topological Charges over
the Fuzzy Sphere, {\tt math-ph/0103003}.


\bibitem{ns} N. A. Nekrasov and A. Schwarz, \cmp {198}{1998}{689};
M. Berkooz, \pl {430}{1998}{237}.

\bibitem{KLY} K.-Y. Kim, B.-H. Lee, and H. S. Yang, Comments
on Instantons on Noncommutative $\r^4$, \hep{0003093}.


\bibitem{J} D. A. Varshalovich, A. N. Moskalev, and V. K.
Khersonskii, {\it Quantum Theory of Angular Momentum} (World
Scientific, Singapore, 1988).


\end{thebibliography}
\end{document}